\documentclass[12pt]{iopart}

\usepackage{cite}
\usepackage{graphicx}  % needed for figures
\usepackage{subcaption}

\begin{document}

\title[Nonlinearity-induced corner states in a kagome lattice]{Nonlinearity-induced corner states in a kagome lattice}

\author{K Prabith$^1$, Georgios Theocharis$^2$ and Rajesh Chaunsali$^1$}
\address{$^1$ Department of Aerospace Engineering, Indian Institute of Science, Bangalore 560012, India}
\address{$^2$ LAUM, CNRS-UMR 6613, Le Mans Universit\'{e}, Avenue Olivier Messiaen, 72085 Le Mans, France}
\ead{rchaunsali@iisc.ac.in}
\vspace{10pt}
\begin{indented}
\item[]April 2025
\end{indented}

\begin{abstract}
Nonlinearity provides a powerful mechanism for controlling energy localization in structured dynamical systems. In this study, we investigate the emergence of nonlinearity-induced energy localization at the corners of a kagome lattice model featuring onsite cubic nonlinearity. Employing quench dynamics simulations and nonlinear continuation methods, we analyze the temporal and spectral characteristics of localized states under strong nonlinearity. Our results demonstrate the formation of stable, localized corner states, strikingly, even within the parameter regime corresponding to the topologically trivial phase of the underlying linear system, which normally lacks such boundary modes. Furthermore, we identify distinct families of nonlinearity-induced corner states residing within the semi-infinite spectral gap above the bulk bands in both the trivial and nontrivial phases. Stability analysis and nonlinear continuation reveal they are intrinsic nonlinear solutions, fundamentally distinct from perturbations of linear topological or bulk states. These findings elucidate a robust mechanism for generating localized states via nonlinearity, independent of linear topological protection, and provide answers to fundamental questions about the nature of nonlinear topological phenomena. The ability to create tunable, localized states in various spectral regions offers potential applications in energy harvesting, wave manipulation, and advanced signal processing.
\end{abstract}

\section{Introduction}
Higher-order topological insulators (HOTIs) have emerged as a significant extension of topological phases of matter over the past few years \cite{xie2021higher}. This field has garnered substantial attention by revealing topologically protected states, such as corner and hinge modes, localized at the boundaries of appropriately dimensioned lattices. Initially, the existence of these states was primarily linked to quantized bulk multipole moments, offering a novel perspective on bulk charge distribution and its boundary consequences \cite{benalcazar2017quantized, benalcazar2017electric, schindler2018higher, peterson2018quantized, serra2018observation, imhof2018topolectrical, mittal2019photonic, qi2020acoustic, dutt2020higher}. However, subsequent work demonstrated that such boundary states can also arise in systems with vanishing quadrupole moments, attributing their origin instead to mechanisms like filling anomalies intertwined with crystalline symmetries \cite{benalcazar2019quantization}. This understanding clarified that topological transitions between trivial and nontrivial phases, characterized by the presence or absence of protected boundary states, can be driven by changes in bulk polarization (e.g., via alternating hopping strengths) without necessarily involving quantized higher-order multipole moments. The fundamental concepts of HOTIs have proven broadly applicable, inspiring research across diverse physical platforms including photonics \cite{xie2018second, xie2019visualization, li2020higher}, acoustics \cite{xue2019realization, wei2021higher, wei20213d, zhang2019deep, chen2019corner, zheng2020three}, mechanics \cite{fan2019elastic, an2022second, chen2021corner, ma2023tuning, chen2023topology, zhou2023visualization, wang2020higher}, electrical circuits \cite{zhang2021experimental, song2020realization, yang2024circuit}, and magnetic systems \cite{chen2020universal, ezawa2018magnetic, guo2023magnetic}.

A defining characteristic of HOTIs is the contrast between their topological phases. In the trivial phase, systems typically behave as conventional insulators, lacking localized states at edges or corners \cite{li2020higher}. Conversely, the nontrivial phase hosts these localized boundary states, whose existence is intrinsically linked to the system's bulk topology—often diagnosed by a topological invariant like the bulk polarization \cite{xie2018second, xie2019visualization, li2020higher}—and protected by underlying lattice symmetries. This inherent robustness against perturbations highlights the potential of HOTIs for applications such as robust waveguiding, signal processing, vibration isolation, and energy harvesting.

While the understanding of HOTIs has advanced significantly, research has predominantly focused on the linear regime, exploring aspects like band topology, symmetry protection, and experimental realizations under weak excitation \cite{xue2019acoustic, ezawa2018higher, wang2021elastic, ezawa2018higher1, kempkes2019robust, proctor2021higher, yatsugi2023higher, yang2024topological, wu2020plane, sil2020first, wakao2020higher, wu2020observation, shen2021investigation, herrera2022corner, zheng2022higher, he2023realization, zhang2024higher, guo2009topological, bolens2019topological}. However, inspired by the rich and often counter-intuitive phenomena observed in nonlinear conventional (first-order) topological systems \cite{smirnova2020nonlinear}, exploring the interplay of higher-order topology and nonlinearity is crucial. Initial investigations into this interplay have already revealed intriguing effects, including the formation of self-induced corner states in 2D lattices \cite{zangeneh2019nonlinear}, robust hinge solitons in 3D systems \cite{tao2020hinge}, and their experimental validation in photonic kagome lattices where Kerr nonlinearity enabled dynamic tuning of corner state frequency and stability \cite{kirsch2021nonlinear}. Related nonlinear phenomena have also been explored considering different lattice configurations \cite{hu2021nonlinear}, various types of nonlinearities \cite{prabith2024nonlinear}, and within open, driven systems \cite{zhang2020nonlinear}.

Interestingly, nonlinearity was also shown to induce localized corner states even within topologically trivial phases \cite{kirsch2021nonlinear}, systems devoid of such states linearly, a phenomenon further explored through quench dynamics \cite{ezawa2021nonlinearity} and nonlinearity management \cite{yi2024delocalization}. While providing crucial insights, these observations fundamentally challenge our understanding of topological protection in the nonlinear regime and raise pivotal questions: Are these nonlinearity-induced states merely modifications of underlying linear modes, or do they represent entirely new classes of states born from the nonlinearity itself? Can nonlinearity similarly populate spectral gaps in nontrivial lattices with localized states that lack linear analogues? Furthermore, what determines the spectral signatures and stability properties of these emergent nonlinear topological states?"

To address these fundamental questions, we investigate the dynamics of a nonlinear HOTI based on a kagome lattice, systematically exploring how nonlinearity-induced corner states emerge and behave under strong nonlinearity in both topologically trivial and nontrivial regimes. Our model incorporates onsite cubic nonlinearity, a form common in diverse physical platforms such as mechanical systems (geometric nonlinearity) \cite{remoissenet2013waves}, optics (Kerr effect) \cite{boyd2008nonlinear}, and magnetism (flux interactions) \cite{remoissenet2013waves}. We employ quench dynamics, exciting the lattice corner and analyzing the transient response via Fast Fourier Transform (FFT) to identify the temporal and spectral features of any emergent localized states. Subsequently, nonlinear continuation techniques are used to rigorously investigate the origin, bifurcation, and stability of these states. Given the fundamental nature of the questions and the generality of our model and approach, this work aims to provide broad insights into identifying and characterizing nonlinearity-induced topological states, potentially paving the way for novel applications reliant on reconfigurable topological modes in photonics, phononics, acoustics, and beyond, impacting areas like wave manipulation, energy concentration, and information storage.

\section{System and its linear spectrum}
Our system consists of a finite kagome lattice comprising $N$ unit cells arranged in a triangular geometry, as shown in Fig. \ref{Fig1}a. Each unit cell, depicted in the inset, contains three masses interconnected by springs with stiffness values $k_1$ (colored red) and $k_2$ (colored blue), representing the intracell and intercell stiffnesses, respectively. These stiffnesses are modulated by a scalar parameter $\gamma$, referred to as the stiffness differential, and are expressed as $k_1 = k(1 + \gamma)$ and $k_2 = k(1 - \gamma)$, where $\gamma \in (-1, 1)$. This alternating pattern of strong and weak stiffnesses imparts the characteristics of a HOTI to the lattice. Furthermore, each mass is grounded via additional springs (colored green) that incorporate a linear stiffness $k_0$ and a nonlinear stiffness $k_{nl}$, with the latter introducing cubic nonlinearity into the system dynamics. We assume a single degree of freedom for each mass, representing its out-of-plane displacement. For convenience, the equations of motion for the masses within the unit cell are expressed in nondimensionalized form by introducing two parameters: $\gamma_0 = k_0 / k$ and $\alpha = a^2 k_{nl} / k$. The resulting equations of motion are as follows \cite{prabith2024nonlinear}:

\begin{figure}[t]
    \includegraphics[width=1\linewidth]{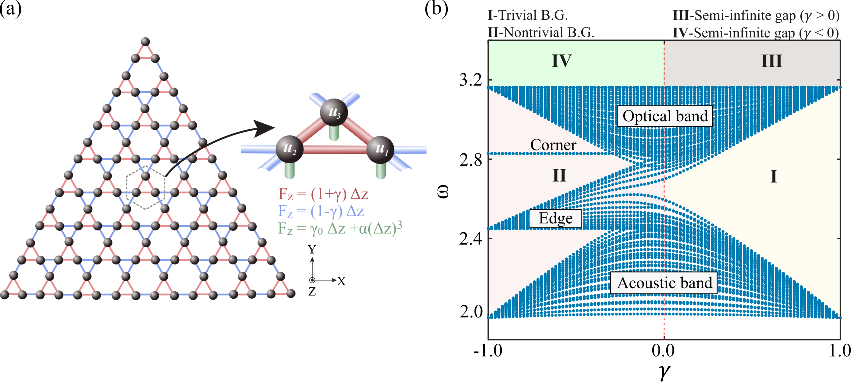}
    \caption{\label{Fig1} A kagome lattice and its linear spectrum: (a) Schematic representation of a kagome lattice comprising point masses interconnected by linear springs (colored red and blue) and attached to the ground through nonlinear springs (colored green). The inset depicts the unit cell in three dimensions. (b) Linear eigenspectrum under fixed boundary conditions, showing the bulk bands—acoustic and optical. The spectrum features a trivial band gap (\textbf{I}) for $\gamma > 0$, a nontrivial topological band gap (\textbf{II}) for $\gamma < 0$, and semi-infinite gaps above the optical band, denoted as \textbf{III} for $\gamma > 0$ and \textbf{IV} for $\gamma < 0$. The topological band gap hosts boundary states, including corner and edge states, as indicated in the spectrum.}
\end{figure}

\begin{eqnarray}
\fl \ddot{u}_{1_{m,n}} + (1+\gamma)(2{u}_{1_{m,n}} - {u}_{2_{m,n}} - {u}_{3_{m,n}})  + (1-\gamma)(2{u}_{1_{m,n}}-{u}_{2_{m+1,n-1}} - {u}_{3_{m,n-1}}) + \gamma_0 {u}_{1_{m,n}}\nonumber \\ 
   + \alpha {u}^3_{1_{m,n}} = 0\nonumber\\
\fl \ddot{u}_{2_{m,n}} + (1+\gamma)(2{u}_{2_{m,n}}-{u}_{3_{m,n}} - {u}_{1_{m,n}}) + (1-\gamma)(2{u}_{2_{m,n}} - {u}_{3_{m-1,n}}-{u}_{1_{m-1,n+1}}) + \gamma_0 {u}_{2_{m,n}} \nonumber \\
  + \alpha {u}^3_{2_{m,n}} = 0\nonumber\\
\fl \ddot{u}_{3_{m,n}} + (1+\gamma)(2{u}_{3_{m,n}}-{u}_{1_{m,n}} - {u}_{2_{m,n}}) + (1-\gamma)(2{u}_{3_{m,n}}-{u}_{1_{m,n+1}} - {u}_{2_{m+1,n}}) + \gamma_0 {u}_{3_{m,n}}\nonumber \\ 
  + \alpha {u}^3_{3_{m,n}} = 0\nonumber\\
\end{eqnarray}
where the variables ${u}_{1_{m,n}}$, ${u}_{2_{m,n}}$, and ${u}_{3_{m,n}}$ denote the nondimensionalized out-of-plane displacements of the three masses, scaled by a reference length $a$, with $m$ and $n$ specifying the position of the unit cell within the lattice. The overdots indicate derivatives with respect to nondimensionalized time.

In the linear regime ($\alpha \to 0$), the eigenspectrum of the lattice as a function of $\gamma$ is computed under fixed boundary conditions and presented in Fig. \ref{Fig1}b for $\gamma_0 = 4$. The spectrum consists of two bulk bands, referred to as the acoustic and optical bands, which are separated by a band gap for nonzero stiffness differentials. The band gap that appears between the bulk bands for $\gamma > 0$ is termed the trivial band gap, labeled as region \textbf{I} (shaded light yellow) in Fig. \ref{Fig1}b. In contrast, the gap for $\gamma < 0$ is classified as nontrivial, as it supports topological states, and is marked as region \textbf{II} (shaded light pink). Specifically, the nontrivial gap hosts edge states for $\gamma < 0$  and corner states for $\gamma < -\frac{1}{3}$ \cite{prabith2024nonlinear,ezawa2018higher}. Additionally, a semi-infinite gap exists above the optical band for all values of the stiffness differential, designated as \textbf{III} (shaded light grey) for $\gamma > 0$ and \textbf{IV} (shaded light green) for $\gamma < 0$. Besides these, a lower band gap appears below the acoustic band; however, it is not highlighted in Fig. \ref{Fig1}b, as this study focuses on hardening nonlinearity, which is not expected to support corner states in the lower band gap. Nevertheless, for softening nonlinearity, this band gap may become relevant and should be taken into account. Notably, in the linear regime, no localized states are observed in any of the band gaps except in the nontrivial gap (\textbf{II}), where edge and corner states emerge due to the topological nature of the lattice.

In the following sections, we extend our analysis to examine the effects of strong hardening nonlinearity ($\alpha = 0.8$) on the eigenspectrum of the finite kagome lattice and to investigate the dynamic characteristics of nonlinearity-induced corner states within the trivial band gap (\textbf{I}) and the semi-infinite gaps (\textbf{III} and \textbf{IV}). The characteristics of nonlinear corner states within the nontrivial band gap (\textbf{II}) have been thoroughly investigated in our previous work \cite{prabith2024nonlinear} and are therefore excluded from the present study. Building on the observation of nonlinearity-induced corner states in the HOTI \cite{kirsch2021nonlinear,ezawa2021nonlinearity}, we conduct a comprehensive numerical investigation to gain deeper insights into their behavior.

\begin{figure}[!]
    \includegraphics[width=1\linewidth]{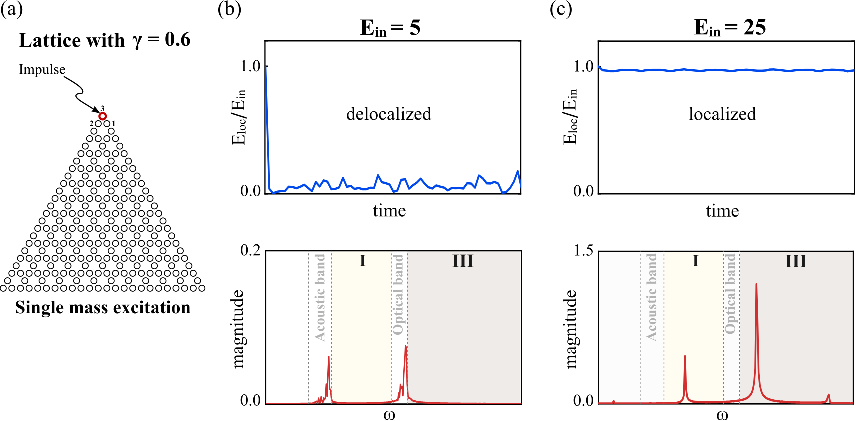}
    \caption{\label{Fig2} Impulse excitation in a lattice with $\gamma > 0$.  (a) Schematic representation of the lattice, showing an initial impulse applied to the top mass ($i = 3$). Fraction of energy localized in the top unit cell ($E_{\mathrm{loc}}/E_{\mathrm{in}}$) and the corresponding FFT spectrum for (b) low input energy ($E_{\mathrm{in}} = 5$) and (c) high input energy ($E_{\mathrm{in}} = 25$). At low input energy, most of the energy leaks into the bulk, and no frequency components appear in the band gap, indicating linear behavior. However, at high input energy, vibrations become confined to the top unit cell, and two dominant frequency components emerge: one in the trivial band gap (\textbf{I}) and the other in the semi-infinite gap (\textbf{III}). Additionally, two smaller peaks appear in the lower band gap and semi-infinite gap at frequencies corresponding to linear combinations of the dominant frequencies.}
\end{figure}

\section{Lattice with $\gamma > 0$}
We begin our analysis with a lattice characterized by a positive stiffness differential ($\gamma = 0.6$), which exhibits a trivial band gap (\textbf{I}) and a semi-infinite gap (\textbf{III}) in the linear limit. To explore signatures of corner localization, we perform quench dynamics by imparting an initial velocity to the corner mass. Upon detecting a localized corner state, we apply numerical continuation to trace the entire solution family and examine its stability characteristics.

\subsection{Quench dynamics}
We apply an impulse to the masses at the top corner through an initial velocity condition, defined as follows:
\begin{equation}
     \dot{u}_i(t) = v_0 \hspace{2cm} \mathrm{at} \hspace{1cm} t = 0
\end{equation}
where $v_0$ represents the initial velocity imparted to the $i$th mass. Consequently, the total energy introduced into the lattice is given by $E_{\mathrm{in}} = \sum \frac{1}{2} \dot{u}_i(0)^2$. Figure \ref{Fig2}a presents a single-mass excitation, where a mass at the top corner of the lattice ($i = 3$) is excited with an impulse. The transient responses of all masses are computed over an extended period ($t = 0.6 \times 10^4$), and the fraction of energy localized in the top unit cell relative to the input energy ($E_{\mathrm{loc}}/E_{\mathrm{in}}$) is evaluated to identify the presence of corner states. Additionally, we determine the frequency content of the excited states by performing a Fourier transform on the displacement of the excited mass.

Figures \ref{Fig2}b and \ref{Fig2}c present $E_{\mathrm{loc}}/E_{\mathrm{in}}$ and the frequency content of the excited mass for low and high input energy, respectively. At low input energy ($E_{\mathrm{in}} = 5$), the impulse applied to the top mass ($i = 3$) leaks entirely into the bulk over time, resulting in negligible $E_{\mathrm{loc}}/E_{\mathrm{in}}$. The corresponding FFT spectrum confirms that only bulk states are excited, indicating the absence of corner states, consistent with the linear lattice spectrum in Fig.~\ref{Fig1}b. However, as the amplitude of the initial velocity increases ($E_{\mathrm{in}} = 25$), nonlinear effects become significant, leading to energy confinement at the top corner of the triangular lattice. This behavior is in agreement with experimental observations reported in Ref.\cite{kirsch2021nonlinear}, where nonlinearity-induced corner states were detected in the trivial phase of a photonic kagome lattice excited by a light pulse at the top corner mass. While Ref.\cite{kirsch2021nonlinear} establishes the existence of such nonlinear corner states, it does not provide insights into their origin, spectral signatures, or stability characteristics. Interestingly, Fig. \ref{Fig2}c reveals the coexistence of two dominant frequency components—one within the trivial band gap (\textbf{I}) and another within the semi-infinite gap (\textbf{III})—suggesting the presence of distinct types of corner states.

\begin{figure}[!t]
    \includegraphics[width=1\linewidth]{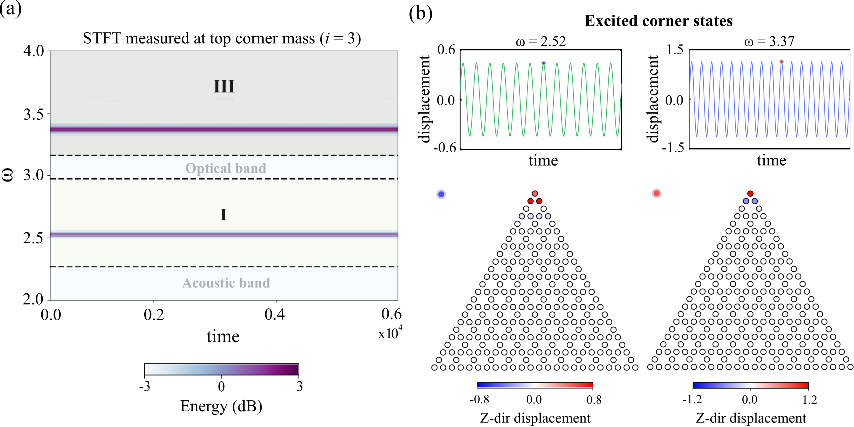}
    \caption{\label{Fig3}  Transient dynamics of the excited corner states during single-mass excitation: (a) STFT of the transient response measured at the top mass ($i = 3$), illustrating that the frequencies of the excited states remain constant over time, indicating their stability. (b) Component signals corresponding to the dominant frequencies ($\omega_1 = 2.52$ and $\omega_2 = 3.37$) and their spatial profiles, extracted from the transient response.}
\end{figure}

To further explore this phenomenon, we compute the short-time Fourier transform (STFT) of the transient response at the top mass ($i = 3$), as depicted in Fig. \ref{Fig3}a. The results indicate that the excited frequencies remain constant over an extended period, suggesting their stable nature. Extracting the component signals corresponding to the dominant frequencies ($\omega_1 = 2.52$ and $\omega_2 = 3.37$) and plotting their spatial profiles in Fig. \ref{Fig3}b reveals distinct corner states: one within the trivial band gap (\textbf{I}), where the three masses in the top unit cell vibrate in phase, and another within the semi-infinite gap (\textbf{III}), where only two masses ($i = 1$ and $i = 2$) in the top unit cell vibrate in phase. This identification of two types of corner states during impulse excitation is significant, as it demonstrates the existence of multiple localized modes in different spectral regions (\textbf{I} and \textbf{III}). Furthermore, the persistence of these states over time suggests that they are stable nonlinear states.

\begin{figure}[!t]
    \includegraphics[width=1\linewidth]{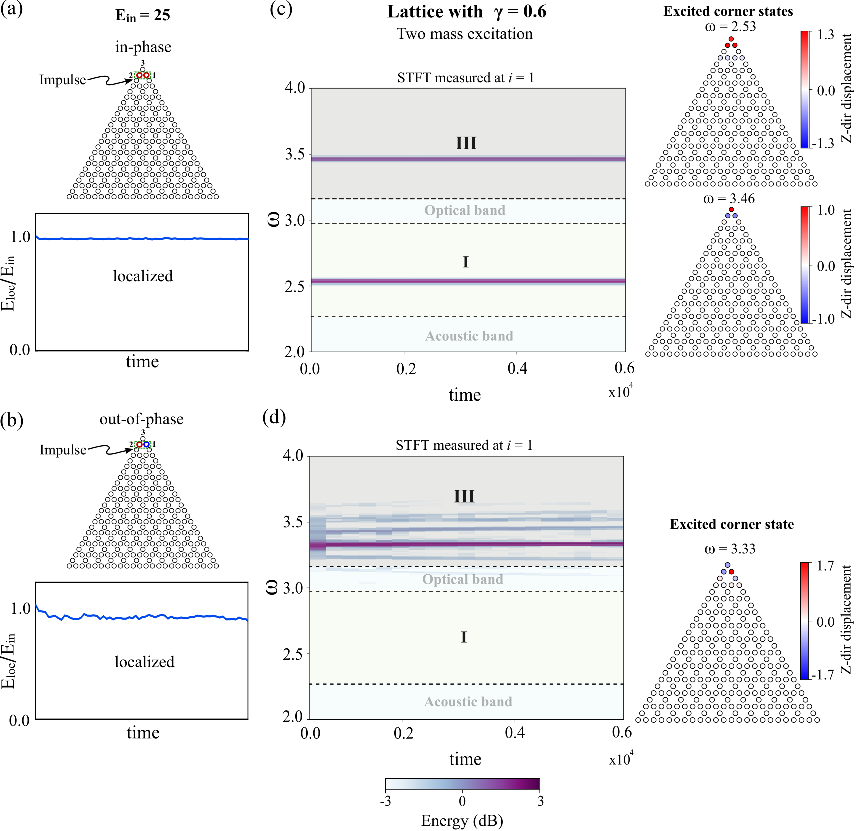}
    \caption{\label{Fig4}Two-mass excitation in a lattice with $\gamma > 0$: (a) \textit{in-phase} excitation of masses at $i = 1$ and $i = 2$. (b) \textit{out-of-phase} excitation of the same masses, showing energy localization at the top unit cell. (c) STFT and spatial profiles corresponding to \textit{in-phase} excitation, illustrating the previously observed corner states. (d) STFT and spatial profiles for \textit{out-of-phase} excitation, revealing a new stable corner state within the semi-infinite gap (\textbf{III}).}
\end{figure}

We also consider other feasible initial conditions, for example, two-mass and three-mass excitations to investigate additional corner states in the nonlinear lattice, and we report only the cases in which new nonlinearity-induced corner states appear. For two-mass excitation, the masses at $i = 1$ and $i = 2$ are excited simultaneously with $E_{\mathrm{in}} = 25$, in both \textit{in-phase} and \textit{out-of-phase} configurations, as shown in Figs. \ref{Fig4}a and \ref{Fig4}b, respectively. In both configurations, energy remains localized at the top unit cell, confirming the presence of corner states. During \textit{in-phase} excitation, the corner states identified from single-mass excitation reappear, as evidenced by the STFT and the corresponding spatial profiles shown in Fig. \ref{Fig4}c. Interestingly, during \textit{out-of-phase} excitation, a new type of corner state emerges within the semi-infinite gap (\textbf{III}), as depicted in Fig. \ref{Fig4}d. This state exhibits a configuration where the two masses ($i = 2$ and $i = 3$) in the top unit cell vibrate \textit{in-phase}, but its orientation differs from that of the state shown in Fig.~\ref{Fig4}c. A mirror image of this state, with masses ($i = 1$ and $i = 3$) vibrating in phase, can also be obtained by reversing the sign of the excitation.

\begin{figure}[t]
    \includegraphics[width=1\linewidth]{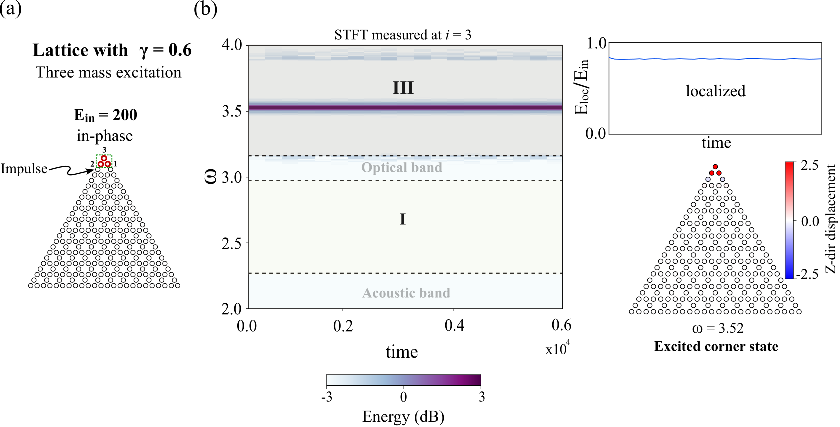}
    \caption{\label{Fig5}Three-mass excitation in a lattice with $\gamma > 0$: (a) \textit{in-phase} excitation of the masses at $i = 1$, $i = 2$, and $i = 3$. (b) The STFT, $E_{\mathrm{loc}}/E_{\mathrm{in}}$, and corresponding spatial profiles demonstrate energy localization within the top unit cell, revealing the emergence of a new stable corner state within the semi-infinite gap (\textbf{III}). This state, characterized by the \textit{in-phase} oscillation of all three masses in the top unit cell, is distinct from the corner state observed within the trivial bandgap in Fig.~\ref{Fig4}c, as evidenced by differences in the vibration patterns of other masses.}
\end{figure}
In the three-mass excitation scenario, we simultaneously excite the masses at sites $i = 1$, $i = 2$, and $i = 3$ with an impulse corresponding to $E_{\mathrm{in}} = 200$, as illustrated in Fig. \ref{Fig5}a. This high-energy excitation is introduced to investigate whether a corner state exists in the semi-infinite gap (\textbf{III}), where the three masses within the top unit cell vibrate in phase. Interestingly, as a consequence of the three-mass excitation, the energy localizes within the top unit cell, and the STFT plot in Fig. \ref{Fig5}b indicates the emergence of a new corner state within the semi-infinite gap (\textbf{III}). The spatial profile shows a configuration in which the three masses of the top unit cell vibrate in phase. It is important to note that this corner state, residing in the semi-infinite gap (\textbf{III}), is distinct from the one found within the trivial band gap (\textbf{I}) in Fig.~\ref{Fig4}c. It oscillates with a large amplitude, and the surrounding masses exhibit a different vibrational response compared to the state in the trivial band gap (\textbf{I}).

Thus far, our quench dynamics analysis of the lattice with $\gamma > 0$ has revealed five distinct corner states: two states, characterized by three masses in the top unit cell vibrating \textit{in-phase}, residing within both the trivial and semi-infinite band gaps; and three states, involving two masses in the top unit cell vibrating \textit{in-phase} (including a mirror image), residing within the semi-infinite gap, as illustrated in Figs.~\ref{Fig4} and \ref{Fig5}. Since these states represent inherent solutions to the nonlinear system, it is necessary to identify their origin. Reference \cite{yi2024delocalization} reported that, in the trivial phase, corner states emerge from the evolution of bulk states through appropriate nonlinearity management, involving a combination of hardening and softening nonlinearities. Although our system employs a simple cubic hardening nonlinearity, it remains necessary to verify whether these nonlinear corner states originate from bulk states or represent isolated solutions. To address this, we systematically trace their solution families through a nonlinear continuation process, using these excited states as initial guesses. The details of this continuation procedure are provided in the following sub-section.

\subsection{Nonlinear continuation}
We employ a Newton solver to trace the family of solutions of the excited corner states by performing a continuation process across various frequencies. This analysis elucidates the origin of nonlinearity-induced corner states by presenting the frequency-lattice energy relationships of the family of excited corner states. The lattice energy of a state is calculated as the sum of the kinetic energy of all masses and the strain energy stored in the intercell, intracell, and grounded springs. Additionally, we assess the linear stability of these states using Floquet theory, with stability determined through the computation of Floquet multipliers. Details of the numerical methods used to compute these states and evaluate their stability are provided in our previous work \cite{prabith2024nonlinear}.

\begin{figure}[!t]
    \includegraphics[width=1\linewidth]{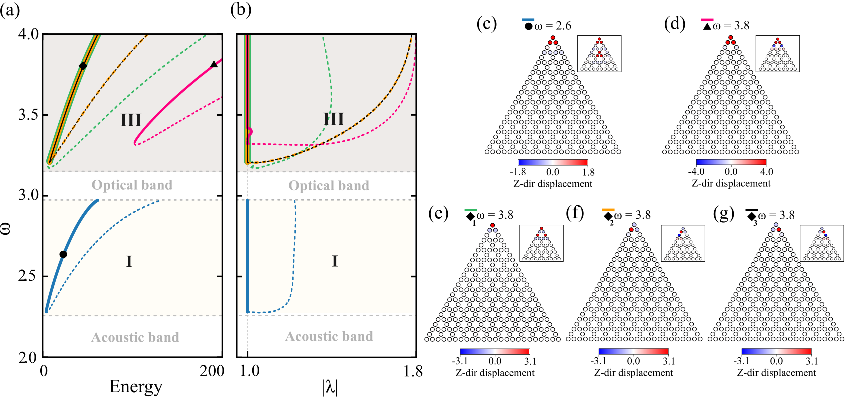}
    \caption{\label{Fig6} Nonlinear continuation of the excited corner states for lattice with $\gamma > 0$: (a) Frequency-lattice energy plot illustrating the evolution of five distinct corner states, represented by blue, green, orange, black and pink curves. Each curve consists of stable (solid lines) and unstable (dashed lines) branches. (b) Maximum amplitude of the Floquet multipliers ($|\lambda|$) used to assess the stability of the corner states. Stability is ensured when $|\lambda| \leq 1$, while instability occurs when $|\lambda| > 1$. (c)-(g) Spatial profiles of the corner states at selected points along the continuation curves, as indicated by circular, triangular and diamond markers in panel (a). The inset depicts the spatial profiles of the corresponding unstable states at the same frequencies.}
\end{figure}
Figure~\ref{Fig6}a displays a frequency-lattice energy plot illustrating the evolution of five different types of corner states, obtained through the continuation procedure. These curves are denoted by different colors and do not originate from any bulk states. Interestingly, they represent isolated solutions and constitute a novel observation in the analysis of a nonlinear HOTI. Each curve has two branches: one stable and the other unstable, represented by solid and dashed lines, respectively.  The stability of each branch is determined by examining the maximum amplitude of the Floquet multipliers ($|\lambda|$), as shown in Fig. \ref{Fig6}b. The solutions remain linearly stable when the maximum amplitude of the Floquet multipliers is less than or equal to unity, while instability arises when it exceeds unity. The spatial profiles of the different stable corner states, indicated by various markers in the continuation curves, are presented in Fig. \ref{Fig6}c–g. These profiles illustrate the diverse ways in which energy can be localized at the corner under identical frequencies but with different excitation conditions. The inset depicts their unstable counterparts at the same frequencies.

In Fig. \ref{Fig6}a, the blue curve represents the family of solutions associated with the first corner state, as indicated by the circular marker in Fig. \ref{Fig6}c, where the three masses in the top unit cell vibrate in phase. These solutions reside within the trivial band gap (\textbf{I}), in agreement with prior observations from the STFT diagrams. As previously noted, the curve exhibits two distinct branches: a stable and an unstable branch, with a transition occurring at $\omega = 2.28$, just above the acoustic band. Notably, the curve does not intersect the acoustic band but instead merges with the optical band at higher energy levels (not shown here).

The pink curve in the semi-infinite gap (\textbf{III}) represents the family of solutions associated with the second corner state, indicated by the triangular marker in Fig. \ref{Fig6}d. In this corner state as well, the three masses in the top unit cell vibrate in phase but with a significantly larger amplitude. Additionally, this state is distinct from the first corner state in terms of the vibrational characteristics of the remaining masses in the system, as observed in both the stable and unstable states. The other three curves, depicted in green, orange, and black, correspond to the remaining three corner states shown in Figs. \ref{Fig6}e–g. These states lie within the semi-infinite gap (\textbf{III}) and are marked with diamond symbols. The corner states shown in Figs. \ref{Fig6}f and \ref{Fig6}g are mirror-symmetric, resulting in equal energies. Consequently, they overlap in Figs. \ref{Fig6}a and \ref{Fig6}b, as indicated by the orange and black curves. All the corner states residing in the semi-infinite gap do not intersect with the optical band and remain localized even at higher energy levels.

In summary, through a combination of quench dynamics and nonlinear continuation, we identify nonlinearity-induced corner states emerging in different spectral regions of the nonlinear HOTI for $\gamma > 0$. These states constitute isolated solutions of the nonlinear system that do not evolve from any bulk states, representing a novel and significant finding in the study of nonlinear HOTIs.

\section{Lattice with $\gamma < 0$}
We now analyze a lattice with a negative stiffness differential ($\gamma = -0.6$), which exhibits a nontrivial band gap (\textbf{II}) and a semi-infinite gap (\textbf{IV}) under linear conditions. During quench dynamics, we do not consider the low impulse single-mass excitation case, as this leads to nonlinear corner states in the nontrivial band gap (\textbf{II}) that have already been discussed in our previous work \cite{prabith2024nonlinear}. Instead, we focus on the two-mass and three-mass excitation with a large impulse, allowing us to explore the emergence of nonlinearity-induced corner states in the semi-infinite gap (\textbf{IV}).

\subsection{Quench dynamics}
\begin{figure}[!t]
    \includegraphics[width=1\linewidth]{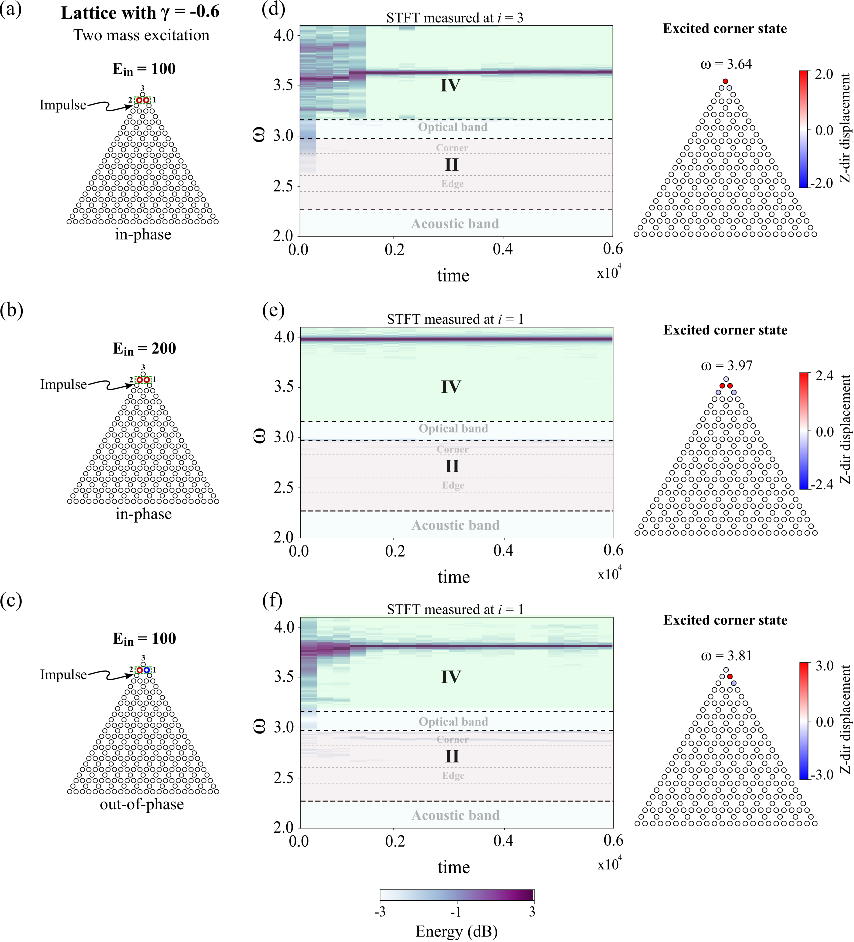}
    \caption{\label{Fig7} Two-mass excitation for a lattice with $\gamma < 0$: (a–c) Schematic representation of two-mass excitations applied at masses $i = 1$ and $i = 2$ simultaneously, considering both \textit{in-phase} and \textit{out-of-phase} conditions. (d–e) STFT diagrams and spatial profiles of the two distinct corner states emerging from \textit{in-phase} excitations with energies $E_{\mathrm{in}} = 100$ and $E_{\mathrm{in}} = 200$. The first corner state (d) exhibits energy localization at the top corner mass ($i = 3$), while the second state (e) shows localization at masses $i = 1$ and $i = 2$ within the top unit cell. (f) STFT diagram and spatial profile of the corner state excited by an \textit{out-of-phase} impulse with $E_{\mathrm{in}} = 100$, revealing a new type of corner state. All these newly observed stable corner states reside within the semi-infinite gap (\textbf{IV}).}
\end{figure}

\begin{figure}[!t]
    \includegraphics[width=1\linewidth]{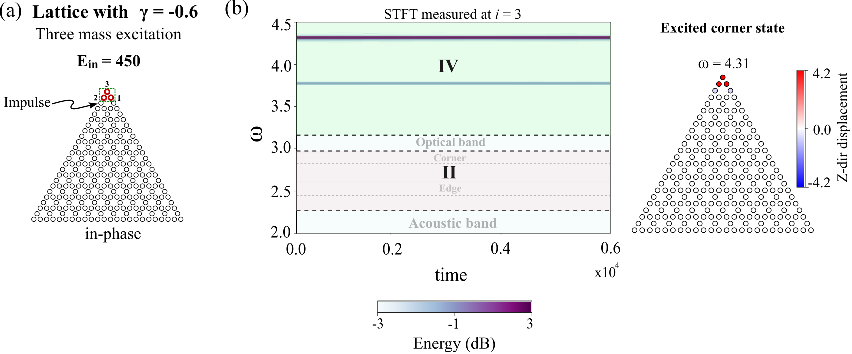}
    \caption{\label{Fig8} Three-mass excitation in a lattice with $\gamma < 0$: (a) Schematic representation of the three-mass excitation, where an in-phase impulse is applied at locations $i = 1$, $i = 2$, and $i = 3$. (b) The STFT diagram and the spatial profile of the excited corner state for an input energy of $E_{\mathrm{in}} = 450$. The results demonstrate the presence of a stable corner state in the semi-infinite gap (\textbf{IV}), where the three masses in the top unit cell vibrate in phase with a higher amplitude.}
\end{figure}

Similar to the analysis conducted for a lattice with $\gamma > 0$, we also perform two-mass and three-mass excitations for the case of $\gamma < 0$, as shown in Figs.~\ref{Fig7} and \ref{Fig8}. We report only the instances in which new nonlinearity-induced corner states emerge.

For two-mass excitations, we apply both \textit{in-phase} and \textit{out-of-phase} excitations at masses $i = 1$ and $i = 2$ simultaneously, as shown in Figs.~\ref{Fig7}a–c. In both cases, excitations of varying energy levels are introduced, leading to the emergence of distinct corner states in the semi-infinite gap (\textbf{IV}). Figures~\ref{Fig7}d–e present the STFT diagrams and the corresponding spatial profiles of the newly excited corner states when \textit{in-phase} excitations with energies $E_{\mathrm{in}} = 100$ and $E_{\mathrm{in}} = 200$ are applied at masses $i = 1$ and $i = 2$ simultaneously. The first corner state, shown in Fig.~\ref{Fig7}d, exhibits a configuration in which most of the energy is localized at the top corner mass of the lattice ($i = 3$), whereas the second corner state in Fig.~\ref{Fig7}e reveals a configuration where the majority of the energy is localized at masses $i = 1$ and $i = 2$ within the top unit cell. On the other hand, for \textit{out-of-phase} excitation, an impulse with energy $E_{\mathrm{in}} = 100$ is applied simultaneously to the masses at $i = 1$ and $i = 2$. The resulting STFT diagram and spatial profile, shown in Fig.~\ref{Fig7}f, indicate the presence of a corner state, despite its resemblance to an edge breather. Notably, the masses at $i = 2$ and $i = 3$ for this state also exhibit small oscillations, reinforcing its classification as a corner state. At higher energy levels, a similar corner state persists; hence, it is not reported in Fig.~\ref{Fig7}. This corner state can also be realized by applying \textit{in-phase} excitation at $i = 2$ and $i = 3$ simultaneously. 

Next, we perform a three-mass excitation by applying an in-phase impulse at locations $i = 1$, $i = 2$, and $i = 3$, as shown in Fig.~\ref{Fig8}a. The STFT diagram and the spatial profile of the excited corner state for an input energy of $E_{\mathrm{in}} = 450$ are presented in Fig.~\ref{Fig8}b. Similar to the case of a lattice with $\gamma > 0$, we observe a stable corner state in the semi-infinite gap (\textbf{IV}), where the three masses in the top unit cell vibrate in phase with a higher amplitude. Its shape indicates that the corner state retains its characteristic profile despite variations in the lattice parameter $\gamma$. By systematically varying the excitation locations within the top unit cell, we identify a total of seven distinct corner states, all of which are employed in the continuation procedure.

\subsection{Nonlinear continuation}
\begin{figure}[!t]
    \includegraphics[width=1\linewidth]{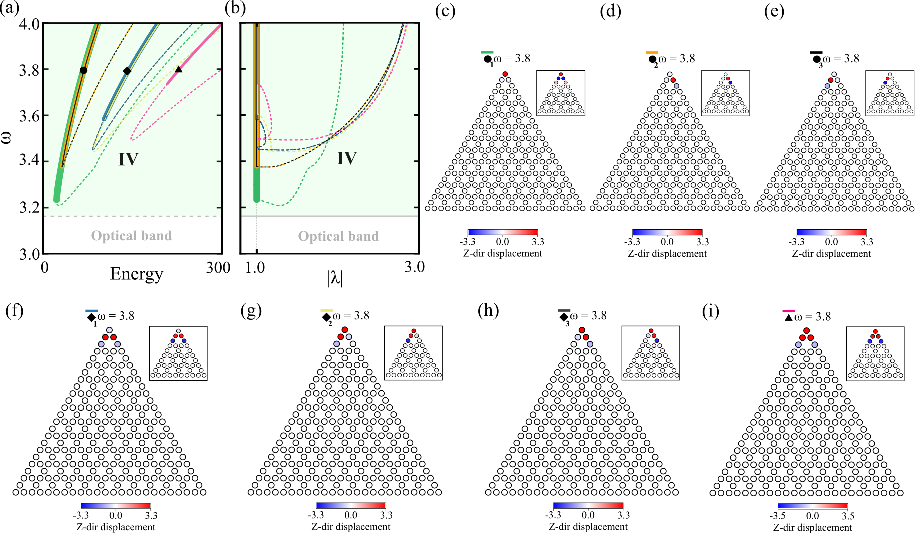}
    \caption{\label{Fig9} Nonlinear continuation of the excited corner states for lattice with $\gamma < 0$: (a) Frequency versus lattice energy, highlighting the emergence of nonlinear corner states. (b) Stability diagram showing stable and unstable branches of the continuation curves. The green, orange, black, blue,  yellow, dark gray, and pink curves correspond to distinct corner states, as depicted in Figs.~\ref{Fig9}c–\ref{Fig9}i. Circular, diamond, and triangular markers indicate the specific corner states presented at \(\omega = 3.8\). The results confirm that these nonlinear corner states are isolated solutions, emerging independently of bulk modes.}
\end{figure}

The continuation procedure is initiated by using the excited corner states as the initial guess for the nonlinear solver. The results of this analysis, including the frequency-lattice energy relationships and the corresponding stability curves, are presented in Figs.~\ref{Fig9}a and \ref{Fig9}b. The green curve represents the family of solutions corresponding to the corner state shown in Fig.~\ref{Fig9}c, characterized by energy localization at the top corner mass (\(i = 3\)). This continuation curve exhibits both stable and unstable branches, with stability switching occurring just above the optical band. These branches extend into the semi-infinite gap as more energy is supplied to the lattice. The orange and black curves correspond to the corner states depicted in Figs.~\ref{Fig9}d and \ref{Fig9}e, which are mirror images of each other and, consequently, overlap in the continuation curves. All these corner states are presented at \(\omega = 3.8\), as indicated by circular markers in Figs.~\ref{Fig9}a.  

The blue curve represents the family of solutions for the corner state shown in Fig.~\ref{Fig9}f, where vibrations are predominantly confined to \(i = 1\) and \(i = 2\). This curve undergoes stability switching far above the optical band and extends into the semi-infinite gap at higher energy levels. The yellow and dark gray curves represent the corner states shown in Figs.~\ref{Fig9}g and \ref{Fig9}h, which are also mirror images of each other. These corner states were obtained by altering the location of the in-phase impulse excitation within the top unit cell compared to that in Fig.~\ref{Fig9}f. All these corner states are presented at \(\omega = 3.8\), as indicated by diamond markers in Fig.~\ref{Fig9}a. Finally, the pink curve in Figs.~\ref{Fig9}a and \ref{Fig9}b represents the family of solutions for the corner state shown in Fig.~\ref{Fig9}i, where all the masses in the top unit cell vibrate in-phase with a higher amplitude. This state is indicated by a triangular marker in Fig.~\ref{Fig9}a, and its mode shape and amplitude distribution closely resemble those observed in the lattice with \(\gamma > 0\) in Fig.~\ref{Fig6}d.  

In summary, the continuation curves presented in Figs.~\ref{Fig9}a and \ref{Fig9}b suggest that these solutions are isolated nonlinear states, emerging independently of any bulk modes. Moreover, a greater number of nonlinearity-induced corner states are observed in the semi-infinite gap (\textbf{IV}) for the lattice with \(\gamma < 0\) compared to the case of \(\gamma > 0\), highlighting the importance of a parameter $\gamma$ in determining the existence and distribution of nonlinear corner states.

\section{Conclusions}
In conclusion, this study presents a novel approach to energy localization in dynamical systems by introducing nonlinearity into a HOTI based on a kagome lattice. Through quench dynamics, we uncover the emergence of nonlinearity-induced corner states that manifest even in the trivial phase and the semi-infinite gap of the trivial as well as nontrivial lattices—regions conventionally regarded as insulating in the linear regime. Our results suggest that these corner states are not merely modifications of existing topological or bulk states. Instead, they emerge as entirely new solutions intrinsic to the nonlinear system, identified through the nonlinear continuation technique. The diversity of corner states observed across different spectral regions provides an efficient mechanism for energy localization, which holds promising applications in areas such as energy harvesting and wave manipulation.

The findings of this study open new avenues for exploring nonlinearity-induced phenomena in wave-based systems, particularly in the context of energy localization at lattice boundaries. Beyond mechanical systems, our results provide valuable insights applicable to a broader range of physical domains, including photonics, acoustics, electrical, and magnetic systems. The existence of corner states, independent of the system’s topological phase, highlights their transformative potential for various applications, such as energy harvesting, optical confinement, topological lasers, acoustic waveguides, noise reduction, sound manipulation, magnonic devices, and spintronic technologies.

Furthermore, future research could focus on the systematic investigation of nonlinear edge and corner states in driven-dissipative topological lattices \cite{pernet2022gap} or non-Hermitian lattices \cite{ezawa2022nonlinear,many2024insensitive}, particularly in kagome or other HOTI lattice geometries. These studies could further deepen our understanding of nonlinear wave dynamics and expand the practical applications of energy localization in complex systems.

\section*{Acknowledgment}
R.C. gratefully acknowledges support from the Science and Engineering Research Board (SERB), India, under Start-up Research Grant No. SRG/2022/001662. K.P. acknowledges the support of the Institute of Eminence (IoE) IISc Postdoctoral Fellowship.

\section*{References}
\bibliographystyle{iopart-num}
\bibliography{biblio.bib}% Produces the bibliography via BibTeX.

\end{document}